\documentclass[12pt]{article}

\usepackage{amsmath}
\usepackage{amssymb}

\newcommand{\eps}{\varepsilon}
\newcommand{\ga}{\gamma}

\begin{document}

\title{Thermodynamic properties of spherical nuclei
\footnote{Yad. Fiz. {\bf 22}, 947-956 (1975) [Sov. J. Nucl. Phys.
{\bf 22,} No.~5, 494-498 (1975)]}}

\author{ A.M. Kamchatnov\\
{\small\it Institute of Spectroscopy, Russian Academy of Sciences,
Troitsk, Moscow Region, 142190 Russia}
}

\maketitle

\begin{abstract}
The effect of the residual interaction between the nucleons (quasiparticles)
on the thermodynamic behavior of spherical nuclei is considered. Thermodynamic
relations are derived for the simplest model of the residual interaction that
has the necessary macroscopic structure previously established.
The range of applicability of the theory is discussed.
\end{abstract}

\section{Introduction}

The nuclear physics data accumulated up to now indicate that there is an essential
difference between spherical and nonspherical nuclei. This difference does not
reduce merely to ``nuclear shape effects" as such, but intimately involves the
internal structure of the nucleus. The shell oscillations of nuclear masses [1]
can serve as an example of the phenomena associated with the difference between
the structures of spherical and nonspherical nuclei. The shell oscillations are
due to the ordered grouping of one-quasiparticle levels in spherical nuclei that
arises because of the existence of the orbital angular momentum quantum number $l$
of the quasiparticle [1,2]. There is no such ordered grouping in nonspherical
nuclei, and the trend of the mass of such nuclei as a function of the number of
nucleons is entirely different. This difference is manifested on a plot of nuclear
masses versus number of nucleons as a break in the curve corresponding to a phase
transformation from the normal phase (nonspherical nuclei) to the ``magic phase"
(spherical nuclei) [3].

The difference between the internal structures of spherical and nonspherical nuclei
does not affect only the ground states: the thermodynamic properties of the two
phases are also different. This difference is manifested, for example, in the
dependence of the form of the spectra of the $\ga$-ray cascades emitted by heated
nuclei on phase of the compound nucleus [4,5]. The energy spectrum of a nonspherical
nucleus is not distorted by the residual interaction and is essentially of the
Fermi-liquid type. Spherical nuclei, on the other hand, apparently cannot exist
without the residual interaction (this was proved in [6] for a very simple nuclear
model), and this interaction has an important effect on their thermodynamic behavior.
As was explained in [2], the residual interaction must have a certain macroscopic
structure, the only interactions between the quasiparticles that are compatible
with the observed pattern of the shell oscillations of nuclear masses being those
that lead to a quadratic dependence of the width of the diffuse region of the Fermi
distribution on the orbital angular momentum $l$ of the quasiparticle:
\begin{equation}\label{1}
    \delta\eps\propto \mathbf{l}^2\cong(l+1/2)^2.
\end{equation}

In this paper we consider the thermodynamics of spherical nuclei on the basis of
the simplest model of the residual interaction that has the required macroscopic
structure.

\section{The Racah-Mottelson model}

The Racah-Mottelson model was used in [2] to describe the residual interaction:
the Hamiltonian for the interaction between quasiparticles on the same $j$ level
is given by
\begin{equation}\label{2}
    H_{int}^f=-G_j\sum_{m,m'>0}a_{m'}^\dagger a_{-m'}^\dagger a_{-m}a_m,
\end{equation}
where $a_m^\dagger$ and $a_m$ are the creation and annihilation operators for a
quasiparticle whose angular momentum has the $z$ projection $m$. The eigenvalues of
this Hamiltonian (see, e.g., [7]) are
\begin{equation}\label{3}
    E_{int}^f=-G_jb_j(\Omega_j-b_j-s_j+1),
\end{equation}
where $2\Omega_j = 2j + 1$, $b_j$ is the number of interacting pairs on level $j$,
and $s_j$ is the seniority quantum number.

To construct the thermodynamics of spherical nuclei, we need a formula for the degree
of degeneracy $\Gamma(s, \Omega)$ (we shall drop the subscript $j$ in the rest of this
section) of the Racah-Mottelson energy levels (3). The symmetry of the zero-seniority
ground-state wave function (see, e.g., [8]) is such that the wave function can be
constructed from single-particle wave functions in only one way. Hence the $s = 0$
ground state is not degenerate: $\Gamma(0, \Omega) = 1$. One more quasiparticle can
be added to the $s = 0$ state in $C_{2\Omega}^1$ ways ($C_{2\Omega}^s = (2\Omega)!/s!
(2\Omega - s)!$ is the number of combinations of $2\Omega$ things taken $s$ at a time),
and we thereby obtain an $s = 1$ state, so that $\Gamma(1, \Omega) = C_{2\Omega}^1$.
Two quasiparticles can be added to the $s = 0$ state with $2b$ quasiparticles in
$C_{2\Omega}^2$ ways, but among them there is one that describes the ground state of $2(b + 1)$
quasiparticles, so that $\Gamma(2, \Omega) = C_{2\Omega}^2-C_{2\Omega}^0$. Arguing
further in the same way we can prove the following general formula for
the degree of degeneracy of the Racah-Mottelson energy levels by mathematical induction:
\begin{equation}\label{4}
    \Gamma(s, \Omega) = C_{2\Omega}^s-C_{2\Omega}^{s-2},\quad 0\leq s\leq\Omega.
\end{equation}

For the case $\Omega\gg1$, $s\gg 1$, we can use Stirling's formula $n! \cong
\sqrt{2\pi n}\exp(n\ln n-n)$ to rewrite Eq.~(4) in the form
\begin{equation}\label{5}
    \Gamma(s,\Omega)\cong\frac{4\Omega(\Omega-s)}{(2\Omega-s)^2}
    \sqrt{\frac\Omega{\pi s(2\Omega-s)}}\exp[(2\Omega)\ln(2\Omega)-s\ln s
    -(2\Omega-s)\ln(2\Omega-s)].
\end{equation}

Now let us find the number $\Gamma(M, s, \Omega)$ of states with seniority $s$
and angular momentum projection $M$. It is clear that the sum of
$\Gamma(M, s, \Omega)$ over all possible values of $M$ is $\Gamma(s, \Omega)$.
In order to find a way of evaluating $\Gamma(M, s, \Omega)$ let us consider the
matrix of Hamiltonian (2) in more detail. The dimension of this matrix is the
number $C_{2\Omega}^N$ of ways in which $N$ fermions can be distributed among
$2\Omega$ cells.  (It is easily seen that the sum of $\Gamma(s,\Omega)$ over all
seniorities $s$ possible for the given $N$ is $C_{2\Omega}^N$.) The Hamiltonian
matrix for the $C_{2\Omega}^N$ basis functions
breaks up into blocks corresponding to different values of the quantum number $s$.
However, the block for the states with a given seniority breaks up in turn into
still smaller blocks, since the Hamiltonian conserves the angular momentum
projection $M$, so that different values
correspond to different blocks. Finally, the block for states with given fixed
values of $s$ and $M$ also consists of smaller blocks, since different sets of
unpaired particles can give rise to the same angular momentum projection $M$.
(Here the term ``pairing" is used in a purely kinematic sense, namely, as an
expression of the fact that two quasiparticles lie in cells corresponding to the
same absolute value but opposite signs of the $z$ projection of the angular momentum.)
For definiteness let is consider the case of even $s$ (and hence even $N$). Then
the ``smallest" blocks in the matrix for Hamiltonian (2) describe states in which
there occur some number $2k$ of unpaired quasiparticles giving a total angular
momentum projection $M$, and $N - 2k$ paired quasiparticles, which have $M = 0$
and seniority $s - 2k$. The number of such blocks is equal to the number
$C(M, 2k, \Omega)$ of ways in which the angular momentum projection $M$ can be
expressed as the sum of $2k$ single-particle projections taken from the set
$\{-j, -j + 1, \ldots, j - 1, j\}$ (no two single-particle projections equal in
modulus but opposite in sign occurring in the sum), while the dimension of each
block is equal to the degree of degeneracy $\Gamma_0(s - 2k, \Omega - 2k)$ of a
state of paired quasiparticles that are distributed among $\Omega - 2k$ free
cells and have seniority $s - 2k$. Thus, the sum of the dimensions of all the
blocks for states with angular momentum projection $M$ and seniority $s$ is
\begin{equation}\label{6}
    \Gamma(M,s,\Omega)=\sum_{k=0}^{s/2}C(M,2k,\Omega)\Gamma_0(s-2k,\Omega-2k).
\end{equation}
A similar expression can also be written for odd $s$.

Direct diagonalization of the Hamiltonian matrix for small values of $\Omega$ and
$N$ leads to values of $\Gamma_0(s, \Omega)$ that can be expressed by the formula
\begin{equation}\label{7}
    \Gamma_0(s,\Omega)=C_{\Omega}^{s/2}-C_{\Omega}^{s/2-1}.
\end{equation}
By definition, $\Gamma_0(s, \Omega)$ is meaningful only for even $s$. If we assume
that formula (7) is valid for all even seniorities $\leq s$, we can prove that it
is valid for seniority $s + 2$ in much the same way as Eq.~(4) was proved: hence
it follows by mathematical induction that (7) is valid for all even seniorities
$0\leq s \leq\Omega$.

Now let us seek the number $C(M, s, n)$ of ways in which $s$ unpaired particles
can give a total angular momentum projection of $M$. Since the formula for
$\Gamma(M, s, \Omega)$ will be used later to describe macroscopic phenomena for
which it is essential that $j\gg 1$, we can set $j\cong\Omega$ (also see [2]),
so that the single-particle angular momentum projections are the elements of the set
$\{-\Omega, -\Omega+1,\ldots,-1, 1, \ldots, \Omega -1, \Omega\}$. Now we shall
construct a generating function for the unknown numbers $C(M. s, \Omega)$.
To do this we consider the sum
\begin{equation}\label{8}
    \sum(\zeta)^{\eps_1}(\zeta^2)^{\eps_2}\ldots(\zeta^\Omega)^{\eps_\Omega}
\end{equation}
taken over the values
\begin{equation}\label{9}
    \eps_1=-1,0,1;\quad \eps_2=-1,0,1;\quad\ldots;\quad \eps_\Omega=-1,0,1
\end{equation}
under the condition that
\begin{equation}\label{10}
    |\eps_1|+|\eps_2|+\ldots+|\eps_\Omega|=s.
\end{equation}
The coefficient of $\zeta^M$ in the sum (8) is $C(M, s, \Omega)$. To take condition
(10) into account, we make use of the identity
\begin{equation}\label{11}
    \frac1{2\pi i}\oint\frac{dz}{z^{|\eps_1|+|\eps_2|+\ldots+|\eps_\Omega|-s+1}}=
    \left\{
    \begin{array}{l}
    0,\quad |\eps_1|+|\eps_2|+\ldots+|\eps_\Omega|\neq s,\\
    1,\quad |\eps_1|+|\eps_2|+\ldots+|\eps_\Omega|=s,
    \end{array}
    \right.
\end{equation}
in which the integration contour goes around the origin. If
we multiply (11) by (8), we can sum over all the values
(9) independently, without regard for condition (10). As a result we obtain the
following expression for the generating function:
\begin{equation}\label{12}
    \sum_MC(M,s,\Omega)\zeta^M=\frac1{2\pi i}\oint\frac{dz}z z^s\prod_{n=1}^\Omega
    \left(1+\frac{\zeta^n+\zeta^{-n}}z\right).
\end{equation}
If we make the substitution $\zeta=e^{it}$ in (12), the sum over $M$ becomes a
finite Fourier series. Using the orthogonality relation
\begin{equation}\label{13}
    \int_{-\pi}^\pi\cos Mt\cos Lt\,dt=\pi\delta_{ML},
\end{equation}
we obtain the following expression for the coefficients:
\begin{equation}\label{14}
    C(M,s,\Omega)=\frac1{(2\pi)^2i}\oint\frac{dz}z z^s\int_{-\pi}^\pi\cos Mt
    \prod_{n=1}^\Omega
    \left(1+\frac2z\cos nt\right).
\end{equation}
Now we expand the product of the expressions in parentheses in powers of $t$ and
extend the integration over the entire real axis, obtaining:
\begin{equation}\label{15}
    C(M,s,\Omega)=\frac1{(2\pi)^2}\sqrt{\frac{3\pi}{\Omega^3}}\oint\frac{dz}z\sqrt{z+2}
    \,z^s\exp\left[\Omega\ln\left(1+\frac2z\right)\right]\exp\left[-\frac{3(z+2)}
    {4\Omega^3}M^2\right].
\end{equation}
This operation is legitimate since the integral over $t$ converges for
$t\sim\sqrt{(z + 2)/\Omega^3}$ and the characteristic values of $z$ (see below) are
such that $z + 2 = 2\Omega/s$, so that $t\sim 1/\Omega\sqrt{s}$. We evaluate the
remaining integral over $z$ by the saddle-point method, choosing the integration
contour so that the convergence of the integral will be due mainly to the function
\begin{equation}\label{16}
    \exp[f(t)]=z^s\exp[\Omega\ln(1+2/z)]=\exp[(s-\Omega)\ln z+\Omega\ln(z+2)].
\end{equation}
At the saddle-point we have
\begin{equation}\label{17}
    z_0=\frac{2(\Omega-s)}s,\quad \left.\frac{d^2f}{dz^2}\right|_{z_0}=
    \frac{s^3}{4\Omega(\Omega-s)},
\end{equation}
and we must therefore integrate over the contour $z = z_0 + iy$. As a result we obtain
\begin{equation}\label{18}
    \begin{split}
    C(M,s,\Omega)=&\sqrt{\frac{3}{2\pi}}\frac1{\Omega\sqrt{s}}\sqrt{\frac\Omega{2\pi s(\Omega-s)}}\\
    &\times\exp[s\ln2+\Omega\ln\Omega-s\ln s-(\Omega-s)\ln(\Omega-s)]\exp(-3M^2/2s\Omega^2).
    \end{split}
\end{equation}
It is easy to show that the condition for the validity of Eq.~(18) is that $s\gg 1$.

To calculate $\Gamma(M, s, \Omega)$ it remains to substitute expressions (7) and (18)
into (6) and perform the summation.  Approximating the right-hand side of (7) with the
aid of Stirling's formula and replacing the summation by integration, we obtain
\begin{equation}\label{19}
    \Gamma(M,s,\Omega)=\sqrt{\frac3{2\pi}^3}\frac{\Omega-s}{2\sqrt{\Omega}}\int_0^{s/2}
    \frac{\exp[\phi(k)]\exp(-3M^2/4k\Omega^2)dk}{k(\Omega-s/2-k)\sqrt{(s/2-k)(\Omega-s/2-k)}},
\end{equation}
where
\begin{equation}\label{20}
    \phi(k)=\Omega\ln\Omega-2k\ln k-(s/2-k)\ln(s/2-k)-(\Omega-s/2-k)\ln(\Omega-s/2-k).
\end{equation}
The function $\phi(k)$ has a maximum at the point
\begin{equation}\label{21}
    k_0=s/2-s^2/4\Omega,
\end{equation}
and at that point
\begin{equation}\label{22}
    \left.(d^2\phi/dk^2)\right|_{k_0}=-16\Omega^3/s^2(2\Omega-s)^2.
\end{equation}
Thus, if
\begin{equation}\label{23}
    s^2(2\Omega-s)^2/\Omega^3\ll s^4/\Omega^2,
\end{equation}
then $\phi(k)$ will have a sharp maximum within the integration range, so we can
replace the other slowly varying functions by their values at $k = k_0$ and integrate
over $k$ from $-\infty$ to $\infty$. As a result we obtain the following final expression:
\begin{equation}\label{24}
\begin{split}
    \Gamma(M,s,\Omega)=&\frac{4\sqrt{3}}\pi\frac{\Omega(\Omega-s)}{s(2\Omega-s)}
    \exp[(2\Omega)\ln(2\Omega)-s\ln s-(2\Omega-s)\ln(2\Omega-s)]\\
    &\times \exp[-3M^2/\Omega s(2\Omega-s)].
    \end{split}
\end{equation}
Condition (23) for the validity of Eq.~(24) is satisfied when $s\gg\sqrt{\Omega}$.
It should be noted, however, that the exponential factors in (24), which are essential
for the thermodynamic applications, are determined only by the position (21) of the
maximum, so that the condition $s\gg\sqrt{\Omega}$ is essentially just a condition
for the validity of the preexponential factor in (24); the exponential factor, on the
other hand, is valid for $s\gg 1$. On integrating (24) over $M$ from $-\infty$ to $\infty$
we again obtain Eq.~(5).

With that we conclude our study of an individual $j$ level of the Racah-Mottelson model
and turn to the application of the model to the nucleus as a whole.

\section{The thermodynamics of spherical nuclei}

Let us calculate the function describing the distribution of quasiparticles in a
nucleus heated to the temperature $T$. The preexponential factor in (5) can be
neglected in calculating the entropy of a single $j$ level with the formula
$S_j =\ln\Gamma(s_j,\Omega_j)$. Then we obtain the following expression for the
entropy of the nucleus as a whole:
\begin{equation}\label{25}
    S=\sum_jS_j=-\sum_j2\Omega_j\left[\frac{s_j}{2\Omega_j}\ln\frac{s_j}{2\Omega_j}
    +\left(1-\frac{s_j}{2\Omega_j}\right)\ln\left(1-\frac{s_j}{2\Omega_j}\right)
    \right].
\end{equation}
To find the function describing the distribution of the quasiparticles among the
levels, we must minimize the free energy
\begin{equation}\label{26}
    F=E-TS,\quad E=\sum_j\left\{(2b_j+s_j)\eps_j-G_jb_j(\Omega_j-b_j-s_j)\right\}
\end{equation}
(here $\eps_j$ is the energy of a quasiparticle on level $j$ without the residual
interaction) under the condition that the number of quasiparticles,
\begin{equation}\label{27}
    \widetilde{N}=\sum_j(2b_j+s_j),
\end{equation}
remains constant. In view of the macroscopic nature of the problem under study we
have neglected the ``1" in Eq.~(3) for the interaction energy (see [2]).

Condition (27) can be easily taken into account by the Lagrange multiplier method.
By varying the function $F' = F - \eps_f \widetilde{N}$ ($\eps_f$ is the chemical
potential) with respect to $b_j$ and $s_j$ we obtain the set of equations
\begin{equation}\label{28}
\begin{split}
    w_j^b&=\frac12-\frac{\eps_j-\eps_f}{G_j\Omega_j}-w_j^s,\\
    w_j^s&=\left[\exp\left(\frac{\eps_j-\eps_f-G_j\Omega_jq_j^b}T\right)+1\right]^{-1},
    \end{split}
\end{equation}
for distribution functions $w_j^b=b_j/\Omega_j$ and $w_j^s=s_j/2\Omega_j$ in the
region $b_j+s_j<\Omega_j,$ $b_j\neq0$ (region II on Fig.~1). On substituting the
expression for $w_j^b$ from the first of Eqs.~(28) into the second we obtain the
following equation for the distribution function $w_j^s$:
\begin{equation}\label{29}
    w_j^s=\left\{\exp\left[\frac{G_j\Omega_j}{2T}(1-2w_j^s)\right]+1\right\}^{-1}
    \quad \text{(region II).}
\end{equation}
On increasing $\eps_j$, we pass from region II to region III, where there are no
paired quasiparticles, i.e., $b_j = 0$. On taking this additional condition into
account in the $s_j$ variation we obtain the following expression for the
distribution function $w_j^s$:
\begin{equation}\label{30}
    w_j^s=\left[\exp\left(\frac{\eps_j-\eps_f}T\right)+1\right]^{-1}
    \quad \text{(region III),}
\end{equation}
i.e., an ordinary Fermi distribution. On decreasing $\eps_j$ we pass from region II
to region I, where all the vacancies permitted by the Pauli principle are filled,
i.e., $b_j + s_j = \Omega_j$. This auxiliary condition gives a Fermi distribution
for the holes:
\begin{equation}\label{31}
    w_j^s=\left[\exp\left(\frac{\eps_f-\eps_j}T\right)+1\right]^{-1}
    \quad \text{(region I).}
\end{equation}
The equations for the boundaries between regions I, II, and III are obtained by
equating the expressions for the distribution functions on the two sides of the
boundary. This gives the equation
\begin{equation}\label{32}
    \frac{\eps_j-\eps_f}T=\frac{G_j\Omega_j}{2T}\tanh\frac{\eps_j-\eps_f}{2T}
\end{equation}
for the boundary between regions II and III. We denote the solution of this equation by
\begin{equation}\label{33}
    \frac{\eps_j-\eps_f}T=\ga\left(\frac{G_j\Omega_j}{2T}\right).
\end{equation}
For the boundary between regions I and II we obtain
\begin{equation}\label{34}
    \frac{\eps_j-\eps_f}T=-\ga\left(\frac{G_j\Omega_j}{2T}\right).
\end{equation}
Since the solution of Eq.~(29) is independent of $\eps_j$, its value within region II
is equal to its value on the boundary (for constant $\Omega_j$) or to the values of
functions (30) and (31) on the corresponding boundaries (33) and (34), so that
\begin{equation}\label{35}
    w_j^s=\left\{\exp\left[ \ga\left(\frac{G_j\Omega_j}{2T}\right)\right]+1\right\}^{-1}
    \quad \text{(region II).}
\end{equation}
A graph of the function $\ga(x)$, defined implicitly by the equation
\begin{equation}\label{36}
    \ga=x\tanh(\ga/2),
\end{equation}
is shown in Fig.~2. We have $\ga(x) \cong\sqrt{6(x-2)}$ as $x \to 2 + 0$, and
$\ga(x) \cong x$ as $x \to\infty$.

Equations (30)-(36) completely determine the distribution function $w_j^s$ and make
it possible to calculate the entropy of a spherical nucleus from Eq.~(25).

It is convenient to do the calculation in the variables $\rho$ and $\widetilde{\beta}$
introduced in [6,2] ($\rho = kR$, $k$ is the wave number of the quasiparticle, and
$R$ is the radius of the nucleus):
\begin{equation}\label{37}
    \begin{split}
    &\rho[\cos\widetilde{\beta}-(\pi/2-\widetilde{\beta})\sin\widetilde{\beta}]=\pi(n+3/4),
    \\ &\widetilde{\beta}=\arcsin(\tilde{l}/\rho),\quad
    \tilde{l}=l+1/2,\qquad dnd\rho=(\rho d\rho/\pi)\cos^2\widetilde{\beta}
    d\widetilde{\beta},
    \end{split}
\end{equation}
where $n$ is the principal quantum number. To the macroscopic accuracy that we are confining
ourselves to, we can replace the total angular momentum $j$ of the quasiparticle by its
orbital angular momentum $l$, so that $\Omega_j\cong l\cong\rho\sin\widetilde{\beta}$.
Taking also into account the fact that in our theory
small values $\widetilde{\beta}$ and $\rho\approx\rho_f$ are important at low
temperatures ($\rho_f = k_fR$, where $k_f$ is the wave number at the Fermi boundary)
we can put
\begin{equation}\label{38}
    \Omega_j\cong \rho_f\widetilde{\beta},\quad G_j=\left.\frac{d\eps}{d\rho}
    \right|_fg_j,\quad T=\left.\frac{d\eps}{d\rho}\right|_f\tau.
\end{equation}
In order to maintain the required macroscopic structure (see the Introduction) in this
same limit, it is necessary, as was shown in [2], that
\begin{equation}\label{39}
    g_j=g\widetilde{\beta}.
\end{equation}
The distribution function $w^s$ assumes the following form in the variables $\rho$ and
$\widetilde{\beta}$ (see Fig.~3):
\begin{equation}\label{40}
    w^s(\rho,\widetilde{\beta})=\left\{
    \begin{array}{l}
    \left[\exp\left(\frac{\rho_f-\rho}\tau\right)+1\right]^{-1},\qquad\qquad \rho-\rho_f
    <-\tau\ga\left(\frac{g\rho_f\widetilde{\beta}^2}{2\tau}\right),\\
    \left[\exp\left(\ga\left(\frac{g\rho_f\widetilde{\beta}^2}{2\tau}\right)\right)+1\right]^{-1},
    \quad  -\tau\ga\left(\frac{g\rho_f\widetilde{\beta}^2}{2\tau}\right)<\rho-\rho_f
    <\tau\ga\left(\frac{g\rho_f\widetilde{\beta}^2}{2\tau}\right),\\
    \left[\exp\left(\frac{\rho-\rho_f}\tau\right)+1\right]^{-1},\qquad\qquad \rho-\rho_f
    >\tau\ga\left(\frac{g\rho_f\widetilde{\beta}^2}{2\tau}\right).
    \end{array}
    \right.
\end{equation}
We also transform to the variables $\rho$ and $\widetilde{\beta}$ in Eq.~(25), and using
the Jacobian $dndl\cong(\rho_f/\pi)d\rho d\widetilde{\beta}$ (see Eq.~(37)) to replace
the summation over the levels by an integration, we obtain
\begin{equation}\label{41}
    S=-\int_{-\infty}^\infty\int_0^\infty2\rho_f\widetilde{\beta}
    \left[w^s\ln w^s+(1-w^s)\ln(1-w^s)\right]
    \frac{\rho_f}\pi d\rho d \widetilde{\beta}.
\end{equation}
Substituting (40) into (41) and integrating, we obtain the following expression for the entropy
of one nucleon component of a spherical nucleus:
\begin{equation}\label{42}
    S=\frac{\pi^3}4\frac{\rho_f\tau^2}g=\frac{\pi^3}4\frac{\rho_f}{g(\left.d\eps/d\rho
    \right|_f)^2}T^2=\frac{\pi^3}4\frac{R^4m^{*2}}{\hbar^4g\rho_f} T^2,
\end{equation}
where $m^* = P_f/v_f$ is the effective mass of a quasiparticle in the nucleus. Taking
into account the fact that nuclear matter is a two-component system, we find
\begin{equation}\label{43}
    S=\frac{b}2\,T^2,\quad b=\frac{\pi^3}2\frac{R^4m^{*2}}{\hbar^4}\left(\frac1{\rho_f^Ng_N}
    +\frac1{\rho_f^Zg_Z}\right).
\end{equation}
Thus, in spherical nuclei (at low enough temperatures---see below) the entropy depends
quadratically on the temperature, in contrast to the frequently assumed linear dependence.
The latter is characteristic of the energy spectrum of a normal Fermi liquid1 [9,10],
but in nuclear physics it can apparently be regarded as fairly reliably established only
for nonspherical nuclei.

The thermodynamic relations
\begin{equation}\label{44}
    E=\frac{b}3T^3,\quad C=\frac{dE}{dT}=bT^2,\quad S=\frac{3^{2/3}}2b^{1/2}E^{2/3},
\end{equation}
follow from Eqs.~(43), so that the level density in a spherical nucleus depends as
follows on the excitation energy:
\begin{equation}\label{45}
    \rho_{l.d.}(E)\propto\exp[S(E)]=\exp\left(\frac{3^{2/3}}2b^{1/2}E^{2/3}\right).
\end{equation}

Now let us calculate the moment of inertia of a spherical nucleus. To do this we
shall examine the Racah-Mottelson states that have not only a definite seniority $s_j$,
but also a fixed angular momentum projection $M_j$. Then the entropy of level $j$
will be $S_j = \ln\Gamma(M_j, s_j,\Omega_j)$,
where the number of states is given by Eq.~(24) (as before, we neglect the
preexponential factor). The distribution functions for the quasiparticles and the
angular momentum $M_j$ are determined from the condition that the free energy
$F' = E - TS - \eps_j\widetilde{N} - \omega M$ be minimum, where $\omega$ is the
angular velocity and $M$ is the total angular momentum projection. Thus, the
$M_j$ dependent terms
\begin{equation}\label{46}
    \sum_j\left[\frac{3M_j^2T}{\Omega_js_j(2\Omega_j-s_j)}-\omega M_j\right]
\end{equation}
are added to the terms determined by formulas (25)--(27). Varying $M_j$ yields
the angular momentum projection
\begin{equation}\label{47}
    M_j=\omega\frac{\Omega_js_j(2\Omega_j-s_j)}{6T}
\end{equation}
of the $j$ level, so that the total angular momentum projection is given by
\begin{equation}\label{48}
    M=\omega\frac2{3T}\sum_j\Omega_j^3w_j^s(1-w_j^s).
\end{equation}
In this case the distribution function $w_j^s$ depends on $\omega$,
but for small $\omega$ we can neglect this dependence in Eq.~(48) and use Eq.~(40).
The coefficient of $\omega$ in (48) is the moment of inertia: in the usual units
($M$ in erg$\cdot$sec and $\omega$ in sec$^{-1}$) it takes the form
\begin{equation}\label{49}
    I=\frac{2\hbar^2}{3T}\sum_j\Omega_j^3w_j^s(1-w_j^s).
\end{equation}
Now we transform (49) to the variables $\rho,$ $\widetilde{\beta}$ and substitute
expression (40) for the distribution function; then after integrating over regions
I--III we obtain
\begin{equation}\label{50}
I=\frac{\pi^3}{12}\frac{\rho_f^3}{g^2}\frac{\hbar^2}{\left.(d\eps/d\rho\right|_f)^3}T^2.
\end{equation}
In our calculation, however, we have not taken into account the effect of the second
functional derivative of the energy with respect to the distribution function [9,10],
which leads to an additional constant factor in Eq.~(50). This factor can be found by
calculating the moment of inertia of a normal Fermi liquid by a method similar to
that used above.  A simple calculation shows that the degree of degeneracy of a state
with $s$ fermions on level $j$ having the angular momentum projection $M$ differs from
(24) only in the preexponential factor, which is not important for thermodynamic
applications. This means that in the thermodynamic sense the seniorities $s$ are
fermion excitations with a dispersion law leading to distribution function (40).
The moment of inertia of a Fermi liquid is therefore given by Eq.~(49) with the
ordinary Fermi distribution substituted for $w^s$.

Now we easily find
\begin{equation}\label{51}
    I_{FL}=\frac4{45\pi}\frac{\hbar^2\rho_f^4}{\left.d\eps/d\rho\right|_f}.
\end{equation}
Here, too, we have neglected the second functional derivative. Requiring that taking
this into account should lead to the rigid-body value $I_{FL} = (2/5)Nm_nR^2 =
(2/5)(2\rho_f^3|/9\pi)m_nR^2$, where $m_n$ is the nucleon mass, we find
\begin{equation}\label{52}
    \left.\frac{d\eps}{d\rho}\right|_f=\frac{\hbar^2\rho_f}{m_nR^2},
\end{equation}
i.e., $\left.d\eps/d\rho\right|_f$ should have the Fermi-gas value. Hence one of
the factors $\left.d\eps/d\rho\right|_f$ in Eq.~(50) should also have the Fermi-gas
value. Taking this into account, as well as the fact that nuclear matter is a
two-component system, we obtain the expression
\begin{equation}\label{53}
    I=\frac{\pi^3}{12}\frac{m^{*2}m_nR^6}{\hbar^4}\left(\frac1{\rho_f^Ng_N}
    +\frac1{\rho_f^Zg_Z}\right)T^2
\end{equation}
for the moment of inertia of a spherical nucleus.

The function describing the distribution of the nuclear levels over the angular
momentum $J$ at fixed temperature has the form (see [11])
\begin{equation}\label{54}
    w_J=\frac{\hbar^2}{IT}\widetilde{J}\exp\left(-\frac{\hbar^2\widetilde{J}^2}
    {2IT}\right),\quad \widetilde{J}=J+\frac12,
\end{equation}
so that the mean and mean square values of the angular momentum are
\begin{equation}\label{55}
    \overline{\widetilde{J}}=\sqrt{\frac{\pi IT}{2\hbar^2}},\quad
    \overline{\widetilde{J}^2}=\frac{2IT}{\hbar^2}.
\end{equation}
Thus we have
\begin{equation}\label{56}
    \overline{\widetilde{J}^2}=\frac4\pi\left({\overline{\widetilde{J}}}\right)^2,
\end{equation}
and this relation can be used to check the correctness and completeness of the
experimental data on level spins. We also note the equation
\begin{equation}\label{57}
    \overline{\widetilde{J}^2}=\frac{m_nR^2}{\hbar^2}\frac{\rho_f^Ng_N^2+\rho_f^Zg_Z^2}
    {g_Ng_Z(\rho_f^Ng_N+\rho_f^Zg_Z)}E,
\end{equation}
which follows from Eqs.~(55), (53), and (44).

\section{Discussion}

As we already mentioned, the formulas derived here are valid only at low temperatures.
In order to find out how low the temperature must be, we take account of the fact that
Eq.~(39) was established only for angles satisfying the condition
\begin{equation}\label{58}
    \widetilde{\beta}\lesssim \frac1{\sqrt{2\rho_f}}
\end{equation}
(see [2]). Hence the quantity
\begin{equation}\label{59}
    \widetilde{\beta}=\sqrt{\frac{4\tau}{g\rho_f}},
\end{equation}
which is characteristic of our theory, marking, as it does, the point at which region
II begins to have a finite extent along the $\eps$ or $\rho$ axis (see Figs.~1 and 3
and Eq.~(40)), should satisfy condition (58). This gives the criterion
\begin{equation}\label{60}
    \tau\ll\frac{g}8\quad\text{or}\quad T\ll\overline{\delta\eps},
\end{equation}
where $\delta\eps= \left.d\eps/d\rho\right|_fg/8$ is the characteristic width of
the diffuse region of the Fermi distribution caused by the residual reaction
(see [2]). The values of $\overline{\delta\eps}$ found in [2] range from 1.9~MeV
(heavy nuclei) to 5.2~MeV (light nuclei) for neutron magic nuclei and from 1.6~MeV
(heavy nuclei) to 3.8~MeV (light nuclei) for proton magic nuclei. However, it is
difficult at present to judge the part played by another criterion,
\begin{equation}\label{61}
    T\ll T_C,
\end{equation}
implied by the theory, where $T_C$ is the Curie temperature at which the spherical
nucleus passes from the magic phase to the normal phase, i.e., becomes nonspherical
(see [3]). We can only assert that condition (61) becomes more important, and may
become dominant, for spherical nuclei that lie close as regards number of nucleons
to the phase transformation point. The $\ga$-ray spectra and radiative widths of
nuclei that are spherical in the ground state confirm this (see [5,12]).

We also note that all the equations we have derived relate the average values of
the thermodynamic quantities to one another, i.e., they are valid only when the
fluctuations are small. However, the temperature fluctuations [10]
\begin{equation}\label{62}
    \Delta T=\frac{T}{\sqrt{C}}=\frac1{\sqrt{b}}
\end{equation}
amount to $\sim0.6$~MeV for $^{208}Pb$, so that the thermodynamic formulas can
only serve as rough estimates at the excitation energies now being investigated
for spherical nuclei. For example, it follows from Eq.~(57) that $\overline{\widetilde{J}^2}
\simeq 4$ for $^{208}$Pb at an excitation energy of 5~MeV, whereas Eqs.~(55)
with the rigid-body moment of inertia and temperature $T\simeq 0.4$~MeV (which
is apparently characteristic of
nonspherical nuclides; see [12]) give $\overline{\widetilde{J}^2}
\simeq 160$.  As well as can be judged from the available data [13] on levels of
$^{208}$Pb, however, the experimental value is $\overline{\widetilde{J}^2}
\simeq 10$, which is in qualitative agreement with the theory under discussion.

Although there is no thermal phase transformation in the present theory, the
macroscopic structure of the residual interaction as a function of the orbital
quantum number $l$ together with the fact (a macroscopic quantum effect) that
the spherical nuclear configuration is unstable in the absence of interaction
between the quasiparticles (see [6]) enables us to suggest a possible reason
for such a transformation. In fact, as the temperature rises region II shifts to
the right on the $l$ axis and more and more quasiparticles fall into regions I and III,
where they conform to the ordinary Fermi distribution. Hence a phase transformation
from the magic phase to the normal phase should take place when there are no longer
enough quasiparticles left in region II to stabilize the spherical shape of the nucleus.

I express my deep gratitude to V.G.~Nosov for numerous discussions and valuable remarks.

\end{document}